\documentstyle[preprint,aps]{revtex}

\begin{document}
\draft
\title{Probabilistic teleportation of two-particle entangled state}
\author{Bao-Sen Shi\thanks{%
E-mail: drshi@ustc.edu.cn}, Yun-Kun Jiang and Guang-Can Guo\thanks{%
E-Mail: gcguo@ustc.edu.cn}}
\address{Lab. of Quantum Communication and Quantum Computation\\
Department of Physics\\
University of Science and Technology of China\\
Hefei, 230026, P.R.China}
\maketitle

\begin{abstract}
In this report, two different probabilistic teleportations of a two-particle
entangled state by pure entangled three-particle state are shown. Their
successful probabilities are different.
\end{abstract}


Quantum teleportation, proposed by Bennett et.al [1], is the process that
transmits an unknown two-state particle, or a qubit from a sender (Alice) to
a receiver (Bob) via a quantum channel with the help of some classical
information. In their scheme, such a quantum channel is represented by a
maximally entangled Bell state. Teleportation has been demonstrated with the
polarization photon [2] and a single coherent mode of field [3] in optical
experiment. Recently, Gorbacher et. al [4] considered the quantum
teleportation of two-particle entangled state by three-particle
entanglement, in their proposal, a three-particle
Greenberger-Horne-Zeilinger [5] (GHZ) state is used as the communication
channel. The three-photon GHZ state has been realized experimentlly [6]. In
the proposal of standard teleportation, a Bell state or a GHZ state is used
to as quantum channel for faithful teleportation. If the entangled state
used as quantum channel is not maximally entangled, the faithful
teleportation is not possible. Although Mor and Horodecki [7] showed that a
`` conclusive '' teleportation of an unknown two-state particle can be made
possible with pure entangled state of two-particle, whether a perfect
teleportation of two-particle entangled state having a finite probability of
success is possible with pure entangled state of three particles or not is
unknown. In this report, a pure entangled three-particle state is
considered, which is used as quantum channel. We present two schemes, by
which, the probabilistic teleportations of two-particle entangled state can
be realized. The successful probabilities of these two schemes are
different. The probability of one scheme is equal to the twice modulus of
the smaller coefficient of the pure entangled three-particle state.

Let a pure entangled three-particle state be in the following state

\begin{equation}
\left| \Phi \right\rangle _{123}=\alpha \left| 000\right\rangle _{123}+\beta
\left| 111\right\rangle _{123}
\end{equation}

where, $\alpha ,\beta $ are reals, $\left| \alpha \right| >\left| \beta
\right| .$ The two-particle entangled state which will be teleported is:

\begin{equation}
\left| \Psi \right\rangle _{45}=a\left| 00\right\rangle _{45}+b\left|
11\right\rangle _{45}
\end{equation}
The particle 3 of the state $\left| \Phi \right\rangle _{123}$ and particle
pair (4, 5) belong to the sender Alice. The state of particle pair (4, 5) is
unknown to Alice. Other two particles 1, 2 belong to receiver Bob. In order
to realize the teleportation, a Bell state measurement on particles 3 and 4
is made by Alice, which will project particles 1, 2 and 5 into the following
state:

\begin{equation}
\left\langle \Phi ^{\pm }\right| _{34}\left. \Phi \right\rangle
_{123}\otimes \left| \Phi \right\rangle _{45}=\frac{\alpha a}{\sqrt{2}}%
\left| 000\right\rangle _{125}\pm \frac{\beta b}{\sqrt{2}}\left|
111\right\rangle _{125}
\end{equation}

\begin{equation}
\left\langle \Psi ^{\pm }\right| _{34}\left. \Phi \right\rangle
_{123}\otimes \left| \Phi \right\rangle _{45}=\frac{\alpha b}{\sqrt{2}}%
\left| 001\right\rangle _{125}\pm \frac{\beta a}{\sqrt{2}}\left|
110\right\rangle _{125}
\end{equation}
where, $\left| \Phi ^{\pm }\right\rangle _{34}=\frac 1{\sqrt{2}}[\left|
00\right\rangle _{34}\pm \left| 11\right\rangle _{34}]$, $\left| \Psi ^{\pm
}\right\rangle _{34}=\frac 1{\sqrt{2}}[\left| 01\right\rangle _{34}\pm
\left| 10\right\rangle _{34}]$. If Eq. (3) is obtained, then a unitary
transformation on particle 5 is made in another basis \{$\left|
x\right\rangle ,\left| y\right\rangle \}$, where the new basis is related to
the old basis \{$\left| 0\right\rangle ,\left| 1\right\rangle \}$ in the
following manner.

\begin{equation}
\left| 0\right\rangle =\alpha \left| x\right\rangle +\beta \left|
y\right\rangle
\end{equation}

\begin{equation}
\left| 1\right\rangle =\beta \left| x\right\rangle -\alpha \left|
y\right\rangle
\end{equation}

which will transform Eq. (3) into the following:

\begin{equation}
\frac 1{\sqrt{2}}[\alpha ^2a\left| 00\right\rangle _{12}\pm \beta ^2b\left|
11\right\rangle _{12}]\left| x\right\rangle _5+\frac{\alpha \beta }{\sqrt{2}}%
[a\left| 00\right\rangle _{12}\mp b\left| 11\right\rangle _{12}]\left|
y\right\rangle _5
\end{equation}

If Eq.(4) is obtained, by the same measurement on particle 5, it will be
transformed into the following:

\begin{equation}
\frac{-1}{\sqrt{2}}[\alpha ^2b\left| 00\right\rangle _{12}\mp \beta
^2a\left| 11\right\rangle _{12}]\left| y\right\rangle _5+\frac{\alpha \beta 
}{\sqrt{2}}[b\left| 00\right\rangle _{12}\pm a\left| 11\right\rangle
_{12}]\left| x\right\rangle _5
\end{equation}
From the above analysis, we can see that if the result of measurement on
particle 5 is $\left| x\right\rangle $ ($\left| y\right\rangle $) in Eq. (7)
(Eq. (8)), particles 1 and 2 are projected the state $\alpha ^2a\left|
00\right\rangle _{12}\pm \beta ^2b\left| 11\right\rangle _{12}$ ( $\alpha
^2b\left| 00\right\rangle _{12}\mp \beta ^2a\left| 11\right\rangle _{12}$),
which can not be rotated back to the desired state without having any
knowledge of the state parameters $a$ and $b$. The states created in Bob's
hands depend not only on $\alpha $ and $\beta ,$ but also on $a$ and $b$.
Unfortunately,  $a$ and $b$ are supposed to be unknown, so it fails to
reproduce the state exactly on Bob's side. That is means if the result of
measurement on particle 5 is $\left| x\right\rangle $ ($\left|
y\right\rangle $) in Eq. (7) (Eq. (8)), the teleportation fails. When   the
result of measurement on particle 5 is $\left| y\right\rangle $ ($\left|
x\right\rangle $) in Eq.(7) (Eq.(8)), if a unitary operator performs on
particles 1 and 2, then a perfect teleportation can be achieved. For
example, if the result of Bell state measurement on particles 3 and 4 is $%
\left| \Phi ^{+}\right\rangle _{34}$, the result of measurement on particle
5 is $\left| y\right\rangle $, then particles 1 and 2 are projected into
state $a\left| 00\right\rangle _{12}-b\left| 11\right\rangle _{12}$. If we
perform $U_{12}=\sigma _{1z}\otimes I_2$ on particles 1 and 2, where $\sigma
_{1z}$ is Pauli operator performed on particle 1, $I_2$ is a unity operator
performed on particle 2, then

\begin{equation}
a\left| 00\right\rangle _{12}-b\left| 11\right\rangle _{12}\stackrel{U_{12}}{%
\rightarrow }a\left| 00\right\rangle _{12}+b\left| 11\right\rangle _{12}
\end{equation}
The right of Eq. (9) is desired state. From the above analysis, we can see
that two measurements on Bob's side are needed. One is a Bell state
measurement on particles 3 and 4 and the other is single particle Von
Neumann measurement on particle 5. The number of classical bits required is
three. Two bits are required for Alice to inform Bob the result of a Bell
measurement which will decide Bob's the unitary operation and the one more
bit is required to tell Bob the result of single particle Von Neumann
measurement which tells Bob whether a successful teleportation is possible.
If a successful teleportation occurs, the unknown state can be reproduced on
Bob's side with fidelity 1. The total probability of the successful
teleportation is $2\left| \alpha ^2\beta ^2\right| .$

Next, we give another scheme, by which, the successful teleportation can be
obtained with probability $2\left| \beta \right| ^2.$

Similiarly, a Bell state measurement is performed by Alice on particles 3
and 4 of the state $\left| \Phi \right\rangle _{123}\otimes \left| \Phi
\right\rangle _{45}$. When Eq. (3) is obtained, a unitary transformation is
made on particle 5. To carry out this evolution, an auxiliary qubit with the
original state $\left| 0\right\rangle _a$ is introduced. Under the basis \{$%
\left| 0\right\rangle _1\left| 0\right\rangle _a$, $\left| 1\right\rangle
_1\left| 0\right\rangle _a$, $\left| 0\right\rangle _1\left| 1\right\rangle
_a$, $\left| 1\right\rangle _1\left| 1\right\rangle _a\}$, a collective
unitary transformation

\begin{equation}
\left[ 
\begin{array}{cccc}
\frac \beta \alpha & 0 & \sqrt{1-\frac{\beta ^2}{\alpha ^2}} & 0 \\ 
0 & 1 & 0 & 0 \\ 
0 & 0 & 0 & -1 \\ 
\sqrt{1-\frac{\beta ^2}{\alpha ^2}} & 0 & -\frac \beta \alpha & 0
\end{array}
\right]
\end{equation}
is made, which will transform Eq.(3) to the result

\begin{equation}
\lbrack \frac{a\beta }{\sqrt{2}}\left| 00\right\rangle _{12}\left|
0\right\rangle _5\pm \frac{\beta b}{\sqrt{2}}\left| 11\right\rangle
_{12}\left| 1\right\rangle _5]\left| 0\right\rangle _a+\frac{\alpha a}{\sqrt{%
2}}\sqrt{1-\frac{\beta ^2}{\alpha ^2}}\left| 00\right\rangle _{12}\left|
0\right\rangle _5\left| 1\right\rangle _a
\end{equation}
A measurement on auxiliary particle follows. The result $\left|
1\right\rangle _a$ means the failed teleportation. If $\left| 0\right\rangle
_a$ is obtained, we make a measurement on particle 5 in another basis \{$%
\left| x\right\rangle ,\left| y\right\rangle \}$, where is related to the
old basis \{$\left| 0\right\rangle ,\left| 1\right\rangle \}$ in the
following manner

\begin{equation}
\left| 0\right\rangle =\frac 1{\sqrt{2}}[\left| x\right\rangle +\left|
y\right\rangle ]
\end{equation}

\begin{equation}
\left| 1\right\rangle =\frac 1{\sqrt{2}}[\left| x\right\rangle -\left|
y\right\rangle ]
\end{equation}
This measurement will project the state of particles 1 and 2 into the
following:

\begin{equation}
\frac \beta 2[a\left| 00\right\rangle _{12}\pm b\left| 11\right\rangle
_{12}]\left| x\right\rangle _5+\frac \beta 2[a\left| 00\right\rangle
_{12}\mp b\left| 11\right\rangle _{12}]\left| y\right\rangle _5
\end{equation}
Obviously, whether the result $\left| x\right\rangle $ or $\left|
y\right\rangle $ is obtained, if a unitary operation is made on particles 1
and 2, a perfect teleportation can be achieved. When Eq. (4) is obtained, it
can be transformed to the result:

\begin{equation}
\lbrack \frac{b\beta }{\sqrt{2}}\left| 00\right\rangle _{12}\left|
0\right\rangle _5\pm \frac{\beta a}{\sqrt{2}}\left| 11\right\rangle
_{12}\left| 1\right\rangle _5]\left| 0\right\rangle _a+\frac{\alpha b}{\sqrt{%
2}}\sqrt{1-\frac{\beta ^2}{\alpha ^2}}\left| 00\right\rangle _{12}\left|
1\right\rangle _5\left| 1\right\rangle _a
\end{equation}
by the unitary transformation

\begin{equation}
\left[ 
\begin{array}{cccc}
0 & \frac \beta \alpha & \sqrt{1-\frac{\beta ^2}{\alpha ^2}} & 0 \\ 
1 & 0 & 0 & 0 \\ 
0 & 0 & 0 & -1 \\ 
0 & \sqrt{1-\frac{\beta ^2}{\alpha ^2}} & -\frac \beta \alpha & 0
\end{array}
\right]
\end{equation}
Similarly, a measurement on auxiliary particle follows. Result $\left|
1\right\rangle _a$ means failed teleportation .When result $\left|
0\right\rangle _a$ is obtained, we also make a measurement on particle 5 in
basis \{$\left| x\right\rangle ,\left| y\right\rangle \}$, then Eq. (15) can
be written as the following:

\begin{equation}
\frac \beta 2[b\left| 00\right\rangle _{12}\pm a\left| 11\right\rangle
_{12}]\left| x\right\rangle _5+\frac \beta 2[b\left| 00\right\rangle
_{12}\mp a\left| 11\right\rangle _{12}]\left| y\right\rangle _5
\end{equation}
Similariy, if a unitary transformation is made on particles 1 and 2, a
perfect teleportation can be achieved. In this scheme, Alice needs to do the
follows for faithful teleportation. Firstly, she needs to make a Bell state
measurement on particles 3 and 4, a unitary operation of two particles 5 and
the auxiliary particle and a single qubit Von-Neumann measurement on
auxiliary particle, then she will get some information by which she can
judge whether the successful teleportation is possible or not. If successful
teleportation is possible, she needs to make another single particle
Von-Neumann measurement on particle 5. The number of classical bits is three
if teleportation is failed or four if successful teleportation is possible.
The total probability of successful teleportation is 2$\beta ^2$, which is
equal to twice modulus of the smaller coefficient of the pure entangled
three-particle state. Contrast to the first scheme, this scheme is more
complicited, it needs more quantum operations and more classical bits if the
teleportation is successful. The advantage of this scheme is the probability
of successful teleportation is higher than that of the first scheme.

In conclusion. In this report, two different probabilistic teleportation of
a two-particle entangled state by pure entangled three-particle state are
shown. The success probability of the first scheme is less than that of the
second scheme. The successful probability of the second scheme is equal to
the twice modulus of the smaller coefficient of the pure entangled
three-particle state. The second scheme is more complicated than the first
scheme.

This subject is supported by the National Natural Science Foundation of
China and the Natioanl Natural Science Foundation for Youth of China.

\end{document}